\documentclass[fleqn,usenatbib]{mnras}
\usepackage[T1]{fontenc}
\usepackage{ae,aecompl}
\usepackage{bm}
\usepackage{graphicx}

\usepackage{amsmath}	
\usepackage{amssymb}	
\usepackage{txfonts}

\title[Origin of polar planets around single stars]{On the origin of polar planets around single stars}

\author[Chen et al.]{Cheng Chen$^{1}$\thanks{Email: c.chen6@leeds.ac.uk}, Stanley A. Baronett$^{2,3}$, C. J. Nixon$^1$ and Rebecca G. Martin$^{2,3}$ 
\\ $^{1}$School of Physics and Astronomy, Sir William Henry Bragg Building, University of Leeds, Leeds LS2 9JT, UK
\\ $^2$Department of Physics and Astronomy,  University of Nevada, Las Vegas, 4505 South Maryland Parkway, Las Vegas, NV 89154, USA 
\\ $^{3}$Nevada Center for Astrophysics, University of Nevada, Las Vegas, 4505 South Maryland Parkway, Las Vegas, NV 89154, USA\\ \vspace{-0.3in}
}

\date{Accepted 2024 June 23. Received 2024 June 22; in original form 2024 March 03}

\pubyear{2024}

\begin{document}
\label{firstpage}
\pagerange{\pageref{firstpage}--\pageref{lastpage}} 
\maketitle


\begin{abstract}
The Rossiter-McLaughlin effect measures the misalignment between a planet's orbital plane and its host star's rotation plane. Around 10$\%$ of planets exhibit misalignments in the approximate range $80 - 125^\circ$, with their origin remaining a mystery. On the other hand, large misalignments may be common in eccentric circumbinary systems due to misaligned discs undergoing polar alignment. If the binary subsequently merges, a polar circumbinary disc -- along with any planets that form within it -- may remain inclined near 90$^{\circ}$ to the merged star's rotation. To test this hypothesis, we present $N$-body simulations of the evolution of a polar circumbinary debris disc  comprised of test particles around an eccentric binary during a binary merger that is induced by tidal dissipation. After the merger, the disc particles remain on near-polar orbits. Interaction of the binary with the polar-aligned gas disc may be required to bring the binary to the small separations that trigger the merger by tides. Our findings imply that planets forming in discs that are polar-aligned to the orbit of a high-eccentricity binary may, following the merger of the binary, provide a possible origin for the population of near-polar planets around single stars.

\end{abstract}


\section{Introduction}
\label{int}


Standard theories of planet formation suggest that planets orbit in the same plane as the rotational plane of the central star because planets are typically expected to form in a planar disc (e.g., the eight planets in the solar system). However, recent analyses of the online database TEPCAT \citep{Southworth2011} and K2-290 \citep{Hjorth2021} have revealed misaligned planets with 3D obliquity $\psi$ \citep[see][for more details on geometry]{Fabrycky2009} that are between $\psi=80^{\circ}-125^{\circ}$ around single stars in 17 out of 156 systems via the Rossiter–-McLaughlin effect during a planet transit \citep{Albrecht2021}. Recently, there are more plausible planets with nearly polar orbits have been found including GJ 3470b, TOI-858Bb and WASP-178b \citep{Stefansson2022, Hagelberg2023, Pagano2024}.

Several mechanisms have been proposed to try to explain misaligned planets around single stars. These mechanisms are summarised in the discussion provided by \cite{Albrecht2021}. They include (1) the von Zeipel--Kozai--Lidov mechanism \citep{vonZeipel1910, Kozai1962, Lidov1962} triggered by an external, massive perturber, which can excite oscillations between the planet's inclination and eccentricity if the external perturber has an inclination above the critical inclination of $\approx$ 39$^\circ$, (2) for high-mass, short-period (hot Jupiter) planets, tidal dissipation can (in some models) result in the obliquity of the planet being near 90$^\circ$ for a period of time \citep{Lai2011, Rogers2013, Anderson2021}, (3) a secular resonance may occur during the late stage of planet formation as the gas disc disperses which can excite the inclination of an inner planet due to the presence of an outer misaligned massive planet \citep{Petrovich2020}, and (4) magnetic warping \citep[e.g.][]{Foucart2011, Lai2011, Romanova2021} could potentially tilt a young protoplanetary disc into an inclined orbit and it is possible for a planet to end up orbiting within such a region of the disc. However, \cite{Albrecht2021} note that there are difficulties for each mechanism in explaining the properties of the observed systems in their sample. An additional possibility is that the disc forms warped due to chaotic infall of gas from the parent star forming region \citep[e.g.][]{Bateetal2010}, and the disc may remain warped for much of its lifetime (e.g. \citealt{Nixon2010}; and much longer than the early timescales on which planets are expected to form; \citealt{Nixon2018,Manara2018,Tychoniec2020}). It is plausible that some, or all, of these mechanisms operate in different systems to produce the observed distribution of misaligned planets.


Here we explore an alternative formation pathway for near-perpendicular planets to form around single stars. Star formation typically proceeds in a chaotic fashion with the resulting stellar systems comprised of multiple stars that can evolve due to capture and exchange encounters \cite[e.g.][]{Bate2018}. The discs of gas that form in and around these stellar systems can arrive with uncorrelated angular momentum direction compared to the stellar system they form in \citep[e.g.][]{Bateetal2010}. Analytical and numerical work has shown that gas discs which are sufficiently misaligned to an eccentric binary precess predominantly around the binary eccentricity vector, leading to ``polar alignment'', in which the gas disc rotates around the binary eccentricity vector and therefore orthogonal to the binary angular momentum vector \citep{Aly2015, MartinandLubow2017}. This process is thought to have created the polar aligned gas discs found in HD 98800 \citep{Kennedy2019}, V773 Tau B \citep{Kenworthy2022}, and the polar aligned debris disc around the binary 99 Herculis \citep[][which presumably arrived in its current orbital configuration before the gas disc dissipated]{Kennedy2012,Smallwood2020}. See also \citet{Lepp2023b,Ceppi2023,Ceppi2024} for recent works on this topic. 

There is growing observational evidence for significant misalignments between the binary orbital plane and the disc plane in young stellar systems \citep[e.g.][]{Czekala2019}. Binaries with short periods (less than around 30 days) are more likely to show aligned discs, while those with longer periods show a broad inclination distribution. This is likely to arise due to a combination of effects, including that the alignment timescale of the disc-binary system grows significantly with increasing binary period and that large misalignments can result in rapid shrinking of the binary orbit \citep[e.g.][]{Nixonetal2011a, Nixonetal2013} which in short period systems (and particularly those with high eccentricity) could result in the merger of the binary.

It therefore seems possible that, in some systems, the binary eccentricity is high enough that the disc polar aligns to the binary orbit and the binary subsequently merges due to a combination of disc-binary interaction and tidal dissipation. Thus, in this letter, we propose that highly-misaligned planets around single stars may attain their orbits while the stellar system is actually a binary (or multiple) star system. Once the central binary merges, the polar planets are left orbiting close to the polar plane, and the final spin of the merged star is essentially in the plane of the original binary orbit (with only a small offset introduced by the initial spins of each star, which need not be aligned to the original binary orbit). 

To test this hypothesis, we present $N$-body simulations of an eccentric binary with a distribution of near-polar planets/planetesimals and follow the orbital evolution of the binary due to the tides the stars induce on each other. We follow the simulation beyond the point at which the stars merge to attain the final configuration of the planets. In Section~\ref{sta}, we describe the setup of the simulation and present our results. Finally, in Section~\ref{dis} and Section~\ref{con}, we present our discussion and conclusions, respectively.

\section{Simulation setup and results}
\label{sta}


We present numerical simulations performed with the publicly available $N$-body simulation code, {\sc rebound}. Two common options for the choice of numerical integrator used by {\sc rebound} are the {\sc whfast} integrator which is a second-order, fixed timestep symplectic Wisdom--Holman integrator with 11th-order symplectic correctors \citep{Rein2015b} and {\sc IAS15} which is a 15th-order Gauss-Radau integrator with variable timesteps \citep{Rein2015a}. For these simulations we have tested both methods and find that they produce very similar results. As such we present results using the IAS15 integrator.

We solve the gravitational equations in the frame of the centre of mass of the binary. We also include the {\sc reboundx} package, an extended library for incorporating additional physics into {\sc rebound}, and we use the constant time lag model for tides between the binary \citep[see][for more details]{Baronett2022}. We note that the dynamics of a gas disc are similar to the particle dynamics except that there is communication between different radii in the gas disc through pressure and viscosity \citep[e.g.][]{Nixon2016}.

The central binary has components of mass $m_1$ and $m_2$ with a total mass of $m_{\rm b}=m_1+m_2$ and a mass ratio of $q_{\rm b}=m_2/m_1$. The binary orbit has  semi-major axis $a_{\rm b}$ and  eccentricity $e_{\rm b}$. We initially consider an equal mass binary with $q_{\rm b}$ = 1.0 with initial eccentricity of $e_{\rm b}$ = 0.8 and the initial argument of periapsis is $\omega_{\rm b}$ = 0.0. We also present results with different binary parameters in Section~\ref{dis}.

\begin{figure*}
    \centering
    \includegraphics[width=7.2cm]{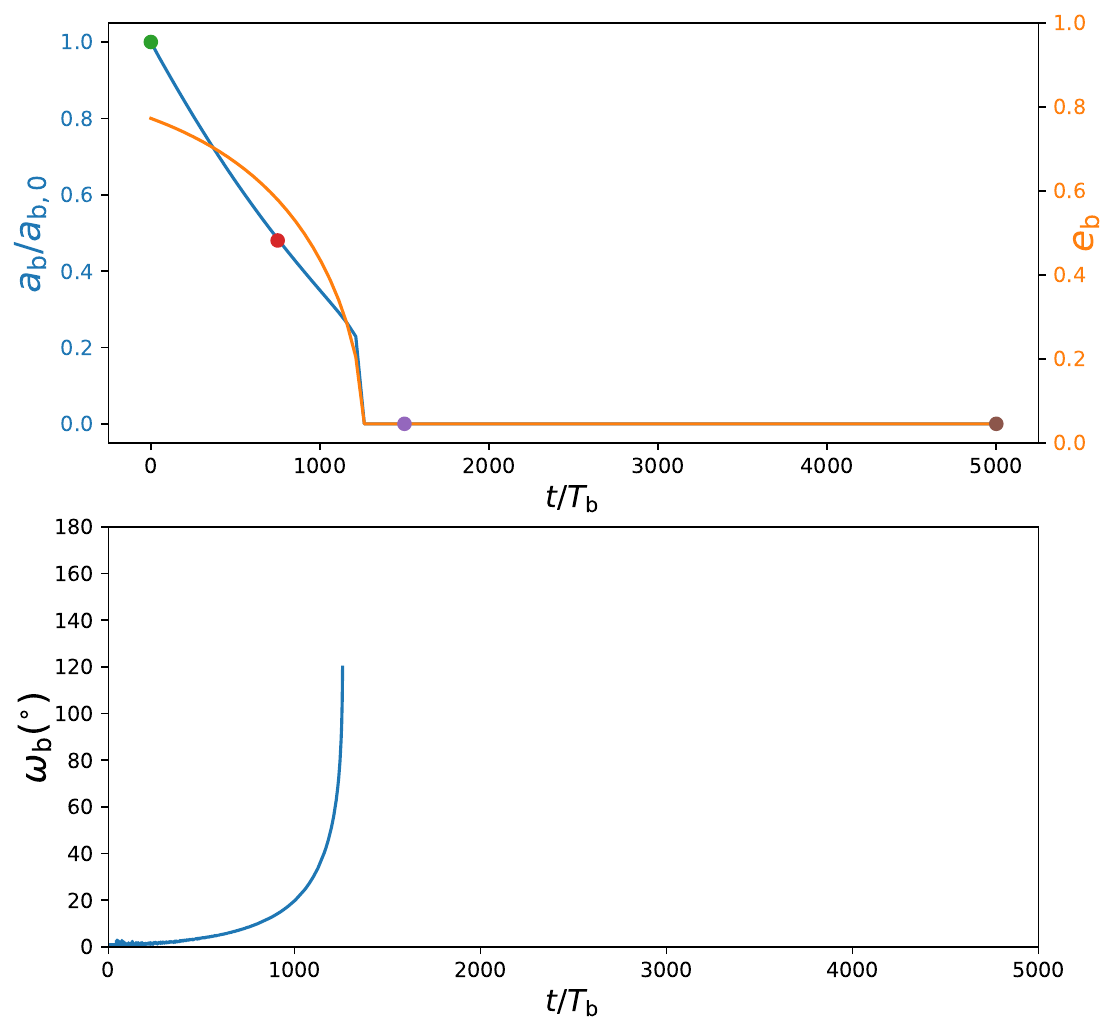}
    \includegraphics[width=9.0cm]{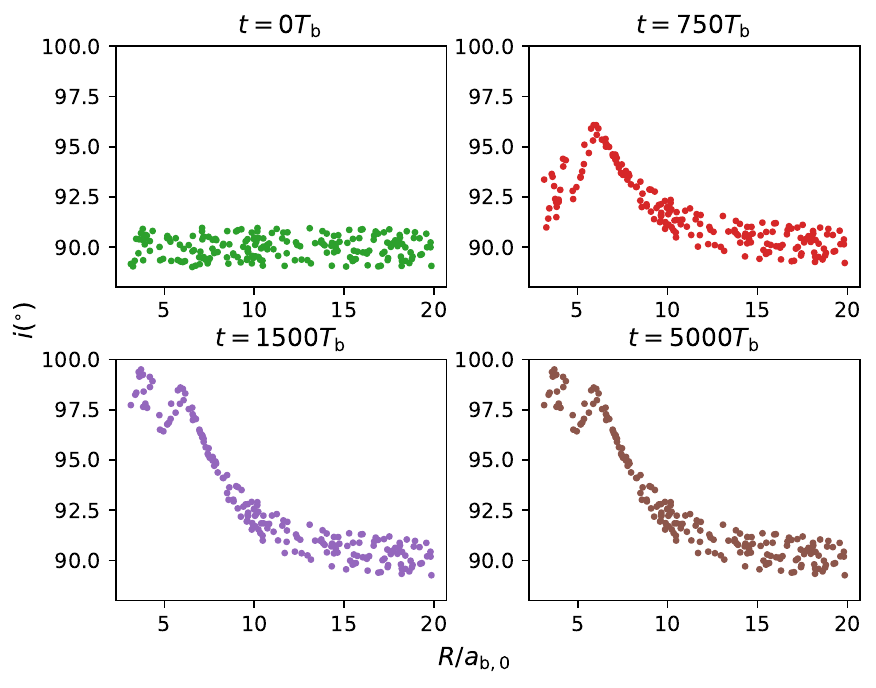}
    \caption{Upper-left panel: the time evolution of $a_{\rm b}$ (blue line) and $e_{\rm b}$ (yellow line).  Lower-left panel: the time evolution of $\omega_{\rm b}$. Right panel: the disc inclination versus the semi-major axis at different times. The four different colours used on the right-hand side represent different times and correspond to the four coloured points in the upper-left panel.
    }   
    \label{fig:merge}
\end{figure*}

A key parameter is the physical radius of the stars, $R_*$,  because the tidal perturbing force is proportional to $R_*^5$ \citep[see equation 8 in][]{Hut1981}. We consider two solar-type main-sequence stars with masses $M_\odot$ and radii $R_{\odot}$.  The initial binary orbital period $T_{\rm b}$ must be long enough that the circumbinary gas disc has sufficient time to align to polar before the binary merges. Observations of main sequence binaries show moderate binary eccentricities for orbital periods $T_{\rm b} \gtrsim 30\,$days and a wide range of $e_{\rm b}$ for $T_{\rm b} \gtrsim 100\,$days \citep[see Fig.~14 in][]{Raghavan2010}. This implies that the merger timescales of moderately eccentric binaries with $T_{\rm b} \gtrsim 30\,$days are longer than the stellar lifetime and the merger will not occur for main sequence binaries with $T_{\rm b} \gtrsim 100\,$days. Thus, we take the initial semi-major axis of the binary to be $a_{\rm b, 0}=0.1\,\rm au$, with initial orbital period $T_{\rm b}= 8.166 \,$days, and thus, $R_{*} = 1 R_{\odot} = 0.05 a_{\rm b,0}$. The upper-left panel of Fig.~\ref{fig:merge} shows the time evolution of the binary semi-major axis, $a_{\rm b}$, (blue line) and the binary eccentricity, $e_{\rm b}$, (orange line). Due to tidal evolution, both $a_{\rm b}$ and $e_{\rm b}$ decrease with time and the binary merges at around $t=1200\,T_{\rm b}$. The lower-left panel shows the evolution of the argument of periapsis. The binary apsidal precession rate increases until the binary merges.

To explore the response of circumbinary planets to the evolution of the binary orbit, we place a disc of 200 test particles on circular orbits around the binary distributed according to a uniform random distribution in the radial range of $3-20\,a_{\rm b,0}$, where $a_{\rm b,0}$ is the initial binary semi-major axis. The inclinations of these particles follow polar orbits ($i=90^{\circ}$), with deviations randomly distributed within $\pm$1.0$^{\circ}$. To set the initial conditions of the test particle orbits we  set the argument of periapsis $\omega_{\rm p}=0$, the true anomaly $\nu_{\rm p}=0$, and the longitude of the ascending nodes measured from the binary semi--major axis to $\phi_{\rm p} =90^{\circ}$. In the absence of binary orbit evolution, the particles remain on stable polar obits according to three-body numerical simulations \citep{Doolin2011, Cuello2019, Chen20201, Childs2021b} and analytical calculations \citep[e.g.][]{Aly2015}.

The upper-left sub-panel on the right side of Fig.~\ref{fig:merge} shows the initial distribution of the inclination of the disc, $i$, versus orbital radius scaled to the initial binary semi-major axis.
The four dots in the top left panel of Fig.~\ref{fig:merge} mark the times at which the plots of inclination $i$ versus orbital radius $R$ are made (shown on the right hand side of Fig.~\ref{fig:merge}). These times are $t = 0$ (green), 750 (red), 1500 (purple) and 5000 (brown) $T_{\rm b}$. Initially, the disc is polar $(i=90^\circ)$ at all orbital radii. After the binary merged around $t=1200\,T_{\rm b}$ ($\approx 26.85\,\rm  yrs$), the particle inclinations no longer vary with time.  As the particles in our simulations are not subject to tides or frictional forces, the innermost edge of the particle disc remains at around $3.0\,a_{\rm b,0}$ and does not shrink with the binary.\footnote{This contrasts with circumbinary gas discs which continue to shrink along with the binary orbit until the decoupling radius at which the binary orbit evolves faster than the disc can move viscously inwards \citep[e.g.][]{Armitage20022}.}
 
A polar aligned test particle around an eccentric binary is in a stationary orbit (meaning there is no nodal precession) with stationary inclination $i_{\rm s}=90^\circ$. Particles that are slightly misaligned to this stationary inclination undergo nodal precession around this inclination.  As a result of the prograde binary apsidal precession, the stationary polar alignment angle increases with the radius of the disc \citep[c.f.][]{Lepp2022, Childs2024}. When the stationary inclination becomes higher than the particle inclination, the particles can undergo nodal precession and the inclination of the particles can get excited. 

\begin{figure*}
    \centering
    \includegraphics[width=17cm]{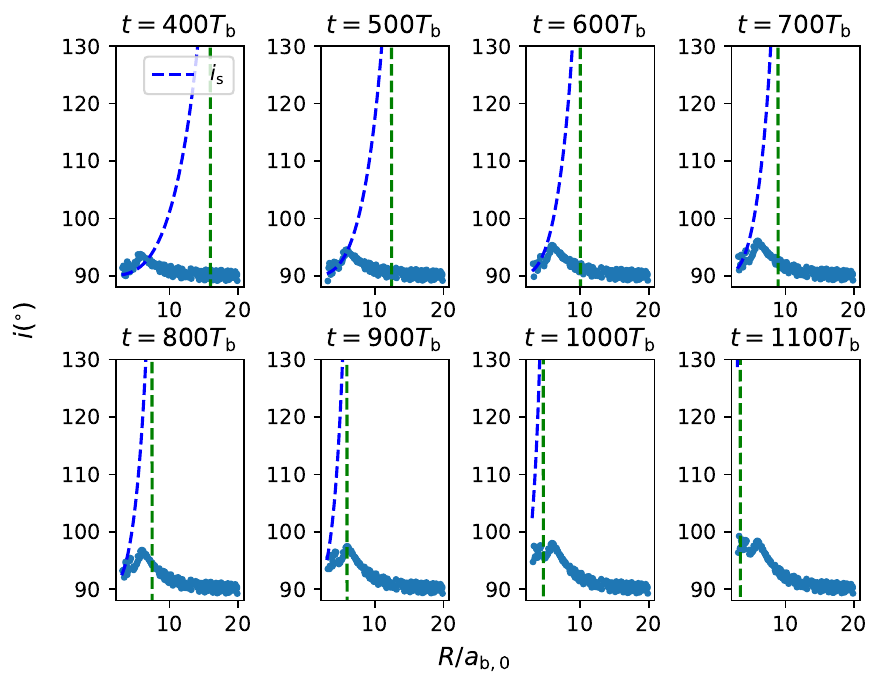}
    \caption{Snapshots of particle inclination $i$ (blue dots) during the binary merger from $t=400 - 1100 T_{\rm b}$ with interval of $100T_{\rm b}$. The blue dashed lines are the stationary inclination $i_{\rm s}$ (equation~\ref{eq}). The green vertical dashed lines indicate where $i_{\rm s}=180^\circ$. The excitation of $i$ only occurs within the green vertical dashed line. }   
    \label{fig:is}
\end{figure*}

We calculate the apsidal precession rate of the binary $\dot{\omega_{\rm b}}$ of the binary and employ a modified form of equation 10 in \citet{Lepp2023}, which gives the stationary inclination, $i_{\rm s}$, as
\begin{equation}
    \cos(i_{\rm s}) = -\dot{\omega_{\rm b}}\times \frac{4}{3
    \sqrt{G}}\frac{(m_1+m_2)^{3/2}}{m_1 m_2}\frac{R^{7/2}}{a_{\rm b}^2}\frac{1}{(1+e_{\rm b}^2)}, 
    \label{eq}
\end{equation}
where $G$ is the gravitational constant.
The stationary inclination is shown by the blue dashed lines in Fig.~\ref{fig:is}. The stationary inclination is initially $i_{\rm s}=90^\circ$ for all radii but it increases during the merger.  Since the particles can nodally precess about this stationary inclination, the magnitude of the inclination oscillations should increase with radius. However, the timescale for these oscillations also increases with radius.  This leads to the two different slopes in the inner and outer parts of the particle disc. The vertical green dashed lines show where the stationary inclination reaches $i_{\rm s}=180^\circ$. For larger particle semi-major axis, the particle no longer nodally precesses about the stationary inclination and instead undergoes nodal circulation with little inclination excitation   \citep[see the lower-right panel of figure 2 in ][]{Lepp2022}.

Scatter points in each panel of Fig.~\ref{fig:is} represent snapshots of the disc inclination $i$ from $t$ = 400 -- 1100 $T_{\rm b}$ with the interval of 100 $T_{\rm b}$. If $|\cos i_{\rm s}|>1$ in Eq.~\ref{eq}, $i$ can not be excited via this oscillation. The dashed green line has moved inward to $R=12a_{\rm b,0}$ at t = 500 $t_{\rm b}$ and the timescale for nodal precession in the outer disc is much longer than the merger timescale, resulting in the outer disc remaining on nearly polar orbits to the end. On the other hand, the middle disc ($R \sim 10 a_{\rm b}$) stops the oscillation earlier than the inner disc while the middle disc has the longer precession timescale than the inner disc. As the result, $i$ of the middle disc only increases a little bit to $i=92^{\circ}$. At the late stage of the binary merging, only the innermost region of the disc still undergoes the oscillation, and the inner disc has a short precession timescale facilitated by the excitation of $i$. Consequently, the inner disc has a higher $i=100^{\circ}$ than the middle and outer disc, leaving a slope structure in two lower-right panels of Fig.~\ref{fig:merge}.

\section{Discussion}
\label{dis}

During the binary merger, the binary undergoes rapid apsidal precession as the stars get closer and closer, inducing the stationary inclination of the disc to increase with increasing semi-major axis. This effect is similar to results when general relativity (GR) within the binary is considered. Recently, \citet{Lepp2022} used the {\sc reboundx} package extension to include GR effects (gr${\_}$full package) and also found the stationary inclination of the test particle increases with increasing semi-major axis. However, when we include the same package in our simulations, the results remain similar to those shown here because the precession timescale of the binary induced by GR is much longer than that induced by tidal interactions for our parameters \citep[see equation 3 in][]{Lepp2022}. Therefore, we can ignore the effect of GR in our scenario, and this is consistent with the $N$-body simulation results in \citet{Antonini2012} that tidal friction is the main mechanism to cause compact stellar binaries and X-ray binaries to merge.

The implementation of tides between the stars in the  \textsc{reboundx} package that we use here employs the constant time lag approximation, which does not evolve the spins of any tidally interacting bodies \citep[][\S~3.1]{Baronett2022}. This implementation is suitable when the spins are expected to be either much greater than the orbital angular frequencies or only negligibly affected by any tidally-mediated angular momentum exchange. 
As \cite{LuReinTamayo2023} have addressed these limitations in a subsequent, separate implementation of self-consistent spin, tidal, and dynamical equations of motion into \textsc{reboundx}, we may use their updated model in a future investigation to assess the impact of stellar spin on the evolution of binary and circumbinary disc (CBD) configurations.

Another possible mechanism to facilitate the binary merger, particularly for wider binary separations, is the interaction of the binary with a circumbinary disc. The transfer of energy and angular momentum through orbital resonances can cause the orbit of the binary to shrink with time \citep[e.g.][]{Artymowicz1994}. However, the orbital evolution of binaries interacting with circumbinary discs is complex, even in the prograde and planar case, and the results can vary depending on the disc and binary properties \citep[e.g.][]{Artymowicz1994, Artymowicz1996,Miranda2017, Tang2017, Munoz2019, Munoz2020, Moody2019, Heath2020}. The main competing effects are the removal of energy and angular momentum by (outer Lindblad) resonances and the increase in binary energy and angular momentum due to accretion of material from the inner edge of the circumbinary disc \citep[see the discussion in][]{Heath2020}. In retrograde discs these two effects both result in the decay of the binary orbit \citep{Nixonetal2011a}.\footnote{While not the focus of this paper, we note that a merger of stars facilitated by interaction with a retrograde circumbinary disc could produce planets with coplanar, retrograde orbits compared to the spin of the central, merged star \citep[cf.][]{Eberle2010}.} While there has not yet been a dedicated study of the orbital evolution of binaries interacting with polar aligned circumbinary discs it is generally found that the binary orbit decays \citep{Aly2015,MartinandLubow2019}. 

Mergers of stars are thought to play a role in explaining the statistics and properties of stars \citep[e.g.][]{Bally2005,Wang2022}. In particular, observations of the stellar multiplicity of solar-type stars show that there is a distinct cut in the period--eccentricity relationship for the 127 binaries at an orbital period of 12 days \citep[see figure 14 in][]{Raghavan2010}. Binaries with orbital period $\lesssim 12\,\rm days$ are close to circular, while those with larger orbital $ 30\,\rm days$ period have a wide range of eccentricities. Additionally, statistical trends suggest that binaries which host polar discs have a mean eccentricity of $e_{\rm b} = 0.65$ \citep{Ceppi2024}.

We can therefore expect that in some regions of parameter space a stellar binary can be brought to merger via interaction with a polar aligned gas disc. As the disc mass and lifetime are finite, this is potentially restricted to binaries with high eccentricity (which is conducive to the formation of polar aligned circumbinary discs) and short orbital periods. The results of, e.g., \cite{Raghavan2010} indicate that there is a lack of binaries with semi-major axes of order 0.1\,au (or less) and eccentricities greater than 0.1. It is reasonable to expect that binaries that form in this region of parameter space may subsequently merge either through tidal effects, interaction with a polar aligned circumbinary disc or a combination of these effects.

\begin{figure*}
    \centering
    \includegraphics[width=0.44\linewidth]{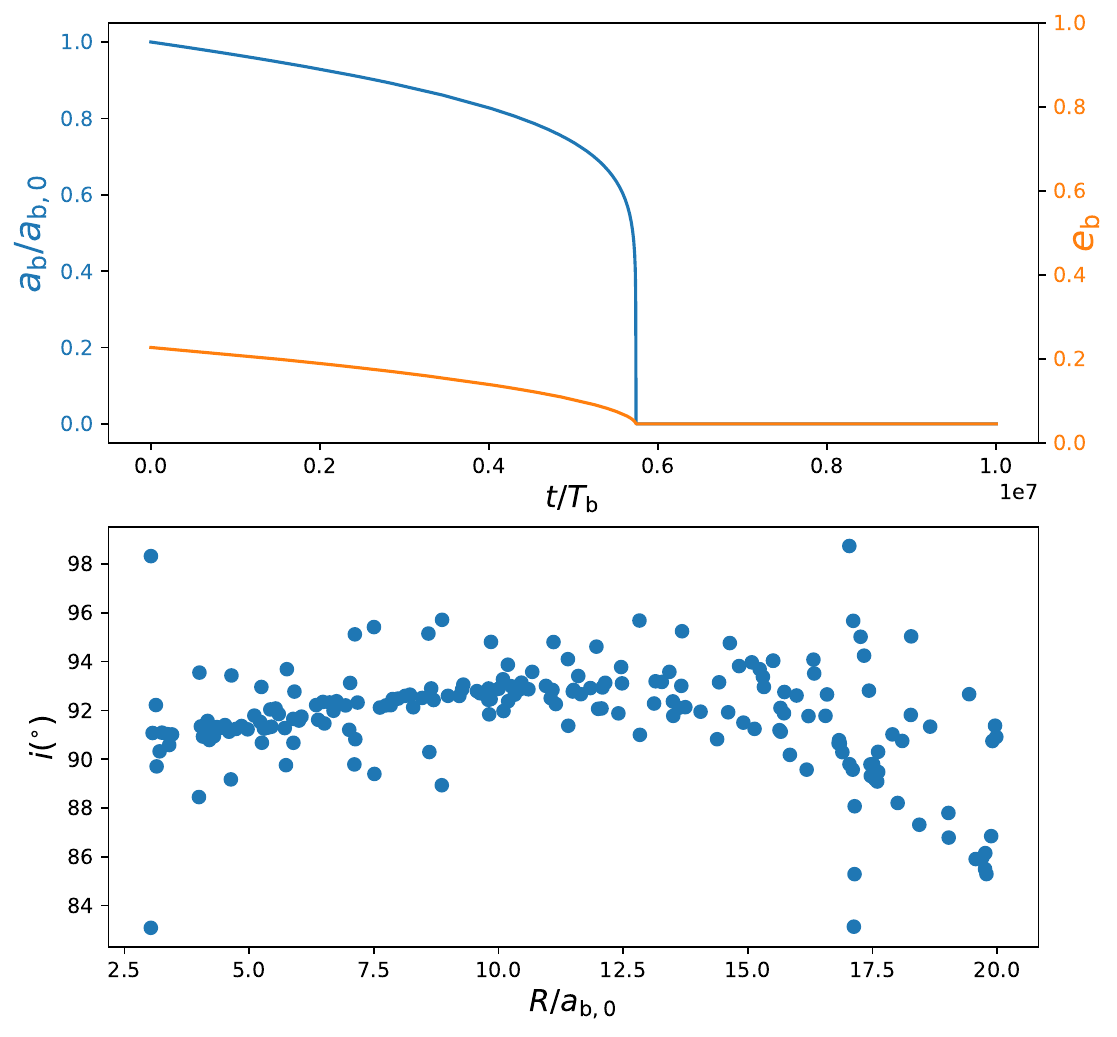}
    \includegraphics[width=0.44\linewidth]{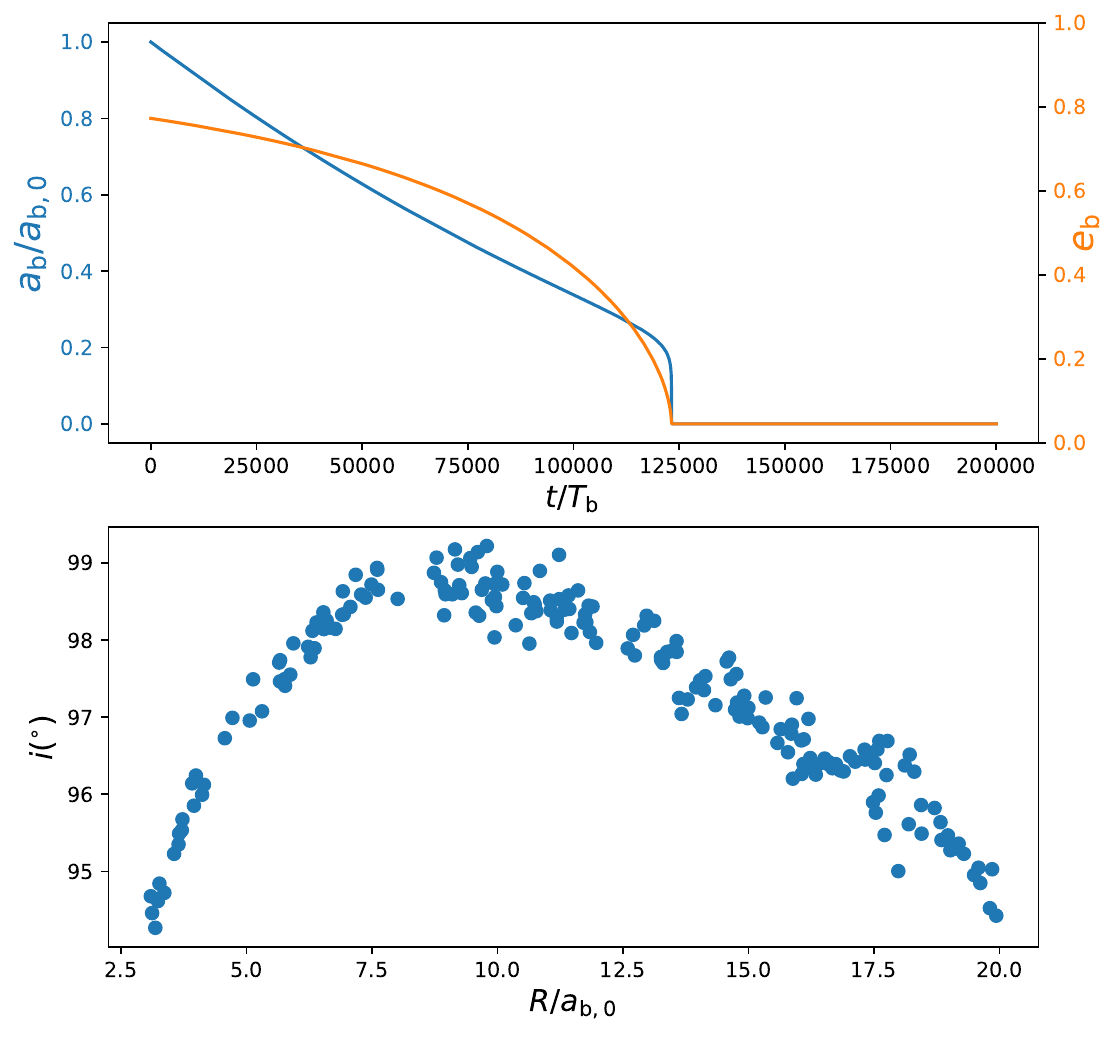}
    \caption{Left panels: The same simulation as Fig.~\ref{fig:merge} except the binary eccentricity is $e_{\rm b}$ = 0.2 and the integration time is longer. Right panels: The same simulation as Fig.~\ref{fig:merge} except $a_{\rm, b0}$ = 0.2 au. The lower panels are snapshots of the particle inclination $i$ at the end of simulations.}
    \label{fig:e065}
\end{figure*}

To examine the effects of varying the initial binary orbital parameters we consider two additional simulations. First, we rerun the same simulation as presented in Section~\ref{sta} except with a lower initial binary eccentricity of $e_{\rm b} = 0.2$. The upper left panel of Fig.~\ref{fig:e065} shows that the binary merges at a time of around 5.9$\times 10^6\,T_{\rm b}$ ($\approx 130,000\,\rm  yrs$). Secondly, we rerun the same simulation as in Section~\ref{sta}, except with a larger initial semi-major axis of $a_{\rm b,0} = 0.2 {\,\rm au}$  (with initial orbital period of $T_{\rm b} = 32.25 \,\rm days$). The upper right panel of Fig.~\ref{fig:e065} shows that in this case the binary merges at a time of around 125,000 $\,T_{\rm b}$ ($\approx 11,000\,\rm  yrs$). Therefore, the merger timescale of a binary with lower $e_{\rm b}$ or larger $a_{\rm b}$ is longer, as expected. The two lower panels show that the particle discs  at the end of simulations are close to polar. There are differences in the distributions with orbital radius due to the different binary apsidal precession rates. The innermost region of the particle disc in the lower-left panel shows  a wide inclination distribution because of the longer merger timescale. Nodal precession  in the innermost region is faster. This effect may explain the wide distribution of inclinations of highly misaligned planets around single stars. 



In addition to apsidal precession, the mass of the third body (or the debris disc), which we do not consider in this study, can also affect $i_{\rm s}$. Analytic solutions and simulations both show that $i_{\rm s}$ depends on $e_{\rm b}$ and on the angular momentum ratio $j$ of the circumbinary disc (or the circumbinary planet) to the binary \citep{MartinandLubow2018b, Lubow2018,Chen20192, MartinandLubow2019} and that $i_{\rm s}$ decreases with increasing $j$ \citep[see equation 17 in][]{MartinandLubow2019}. Thus, a CBD or the circumbinary planet which inherits the high tilt of the disc can start precessing \citep{Cuello2019} and the final inclination after merging can still be lower than 90$^\circ$, and this effect may also explain the wide distribution of inclinations of highly misaligned planets around single stars. On the other hand, the effect of the angular momentum exchange between $e_{\rm b}$ and the inclination of the misaligned third body plays a role if $j$ is large enough. In Figures 3 and 4 of \citet{Chen20192}, simulations show that $e_{\rm b}$ can be excited from 0.2 to above 0.7 if $j$ is large and the mass fraction of the binary is small. Additionally, $e_{\rm b}$ increases with increasing deviation between $i_{\rm s}$ and $i$. Thus, if planet formation occurs while the disc is young, it is likely that planets formed inside the disc could decouple from the disc \citep[e.g.][]{Martin2016,Lubow2016,franchini2020}. A decoupled yet massive circumbinary planet, inheriting the originally high tilt of the disc aligned to $i_{\rm s}$, may tremendously excite $e_{\rm b}$ and speed up the binary merging process even if the binary only has a moderate $e_{\rm b}$ initially.

\section{Conclusion}
\label{con}
We present a new mechanism to form highly misaligned planets around a single star. The process is as follows. First, a misaligned gas disc around an eccentric binary settles into a polar aligned configuration. Second, the binary is driven to merge due to the tides between the two stars if they initially formed close enough or through other effects such as interaction with a circumbinary gas disc.  Third, the formation of planets in the (near-)polar gas disc results in planets with orbits that are near-perpendicular to the spin plane of the merged star which is approximately in the orbital plane of the original binary system.\footnote{Note that the planets may form prior to the binary merger and remain in their near-polar orbits during and after the stellar merger.}

Using $N$-body simulations, we consider the tidal dissipation between two solar-type stars, initially in a binary configuration, surrounded by a polar disc of test particles that represent possible planets with an inclination of about 90$^\circ$. Tidal dissipation between the stars causes the binary to merge, leaving a polar disc around a single star. Moreover, the fast apsidal precession of the binary increases the stationary polar alignment angle of the disc, similar to an effect found with GR \citep[e.g.][]{Lepp2022}. Thus, the disc angle after the merger can be higher than 90$^\circ$. A smaller angle is also possible if the mass of the planet is taken into account \citep[e.g.][]{Chen20192} or if the particle nodal precession timescale is short compared to the merger timescale. Therefore, it is possible that a range of high-inclination planets can be caused by this mechanism. Future simulations that include the gas dynamics of both the disc-binary interaction and the hydrodynamics of the merger process will provide valuable insights into the details of this scenario. 

\section*{Data Availability}
The simulations in this paper can be reproduced by using
the REBOUND code (Astrophysics Source Code Library identifier ascl.net/1110.016). The data underlying this article will be shared on reasonable request to the corresponding author.

\section*{Acknowledgements}
Computer support was provided by UNLV's National Supercomputing Center and DiRAC Data Intensive service at Leicester, operated by the University of Leicester IT Services, which forms part of the STFC DiRAC HPC Facility (www.dirac.ac.uk). CC and CJN acknowledge support from the Science and Technology Facilities Council (grant number ST/Y000544/1). CJN acknowledges support from the Leverhulme Trust (grant number RPG-2021-380). RGM acknowledges support from NASA through grant 80NSSC19K0443. Simulations in this paper made use of the REBOUND code and the REBOUNDx code which can be downloaded freely at http://github.com/hannorein/rebound and https://github.com/dtamayo/reboundx)



\bibliographystyle{mnras}
\bibliography{main} 

\bsp	
\label{lastpage}
\end{document}